\documentclass[prd,showpacs,preprintnumbers,amsmath,amssymb,superscriptaddress]{revtex4}
\usepackage{graphics}
\usepackage{rotating}

\begin{document}

\title{Constraints on the solid dark universe model}
\author{Richard A. Battye}
\affiliation{Jodrell Bank Observatory, School of Physics and
Astronomy, University of Manchester, Macclesfield, Cheshire SK11
9DL, UK}
\author{Adam Moss}
\affiliation{Jodrell Bank Observatory, School of Physics and
Astronomy, University of Manchester, Macclesfield, Cheshire SK11
9DL, UK}
\date{25th February 2005}
\preprint{}
\begin{abstract}

If the dark energy is modelled as a relativistic elastic solid then the standard CDM and $\Lambda$CDM models, as well as lattice configurations of cosmic strings or domain walls, are points in the two-dimensional parameter space $(w,c_{\rm s}^2)$. We present a detailed analysis of the best fitting cosmological parameters in this model using data from a range of observations. We find that the $\chi^2$ is improved by $\sim 10$ by including the two parameters and that the $w=-1$ $\Lambda$CDM model is only the best fit to the data when a large number of different datasets are included. Using CMB observations alone we find that $w=-0.38\pm 0.16 $ and with the addition of Large-Scale Structure data $w=-0.62\pm 0.15 $ and $\log c_{\rm s}=-0.77\pm 0.28$. We conclude that the models based on topological defects provide a good fit to the current data, although $\Lambda$CDM cannot be ruled out.
\end{abstract}
\maketitle

\section{introduction}
\label{intro}

A Standard Cosmological Model has emerged over the past few years which provides an acceptable fit to all the available data~\cite{spergel,tegmark}. It is often known as the $\Lambda$CDM model since the two dominate energy density components of the universe at the present epoch are provided by cold dark matter (CDM) and a cosmological constant, $\Lambda$; their densities relative to the critical density are denoted $\Omega_{\rm c}$ and $\Omega_{\Lambda}$, respectively. The model also includes a smaller fraction of baryons, given by $\Omega_{\rm b}$, and density fluctuations created during an epoch of inflation quantified by an amplitude, $A_{\rm S}$, and spectral index, $n_{\rm S}$. The expansion rate of the universe at the present epoch is given by $H_{0}=100h\,{\rm km}\,{\rm sec}^{-1}\,{\rm Mpc}^{-1}$ and the optical depth to the epoch of reionization is given by $\tau_R$. The age of the Universe, $t_0$, and the redshift of reionization, $z_{\rm re}$, can be computed from these parameters.

Unfortunately, the value of $\Lambda$ required to make this model viable is very much in excess of that which is predicted by standard arguments from particle theory and much, largely unsuccessful, effort as been expended to try and explain the apparent discrepancy between theory and observation. The important feature of $\Lambda$ in the observational context is that it is constant in space and time and therefore does not cluster into virialized objects, and that it violates the weak energy condition, $1+3w>0$ if the pressure and density are related $P=w\rho$, which leads accelerated expansion in the late universe. The generalized concept of dark energy has been introduced to represent any energy component for which $1+3w<0$ and whose Jeans length is sufficiently large to prevent clustering.

A dynamical cosmological constant, often dubbed Quintessence, based on scalar fields has been developed~\cite{PR} as a dynamical alternative to $\Lambda$ and is the most popular dark energy candidate. If a scalar field, $\phi$, rolls slowly down a sufficiently shallow potential then a fluid with $w<-1/3$ and a large Jeans length can be arranged by a suitable choice of potential. However, in order for the dynamics of the scalar field to be super-damped, the Compton wavelength of the field needs to be of order the Hubble radius today which implies that the mass must be $m_{\phi}\sim H_0\sim 10^{-33}{\rm eV}$. Such small values of $m_{\phi}$ are difficult to achieve within particle physics models due to the effects of quantum corrections~\cite{DJ} and such light particles would naturally couple to the fields of the Standard Model of Particle Physics~\cite{carroll}. While it provides an interesting phenomenological model for dark energy, in many ways Quintessence provides no better micro-physical motivation than $\Lambda$. 

The dark energy cannot be a perfect fluid since the sound speed $c_{\rm s}$ is given by\begin{equation}
c_{\rm s}^2={dP\over d\rho}=w,
\end{equation}
which is $<0$ leading to instabilities on the smallest scales. A suggestion~\cite{BS} has been made that the inclusion of rigidity, $\mu$, in the fluid, making it an elastic solid, could provide the necessary stability. By definition a solid not only supports longitudinal (scalar) perturbations, but also transverse (vector) perturbations. The sound speed of these vector modes is given by $c_{\rm v}^2=\mu/(\rho+P)$ and it can be shown that~\cite{B1,B2}
\begin{equation}
c_{\rm s}^2={dP\over d\rho}+{4\over 3}c_{\rm v}^2=w+{4\over 3}c_{\rm v}^2.
\end{equation}
This equation implies that if $\mu/\rho$ is sufficiently large then $c_{\rm s}^2>0$ even if $w<-1/3$. Apart from their role in stabilizing the scalar modes, we will not consider the possible effects of these vector modes here, leaving their detailed exploration to a future publication~\cite{BM2}.

\begin{figure} 
\centering
\mbox{\resizebox{0.5\textwidth}{!}{\includegraphics{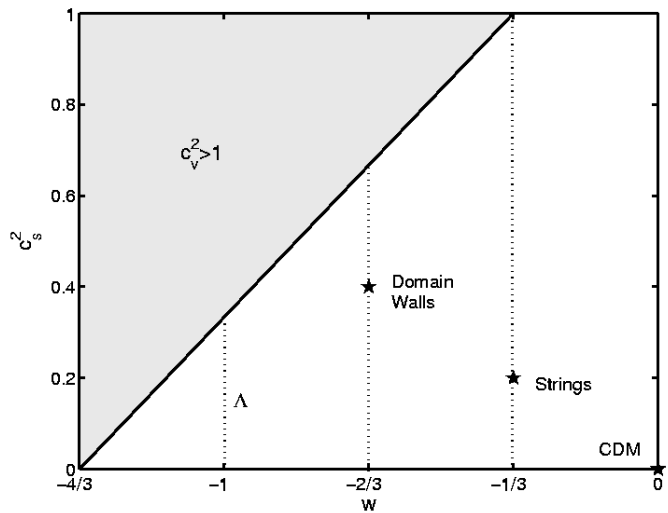}}}
\caption{Allowed parameter space of SDE models. The lines $w=0$, $w=-1/3$, $w=-2/3$ and $w=-1$ correspond to the dark energy models which scale like CDM, static cosmic strings, static domain walls and $\Lambda$ respectively. The point $(w,c_{\rm s}^2)=(0,0)$ corresponds to CDM, $(-1/3,1/5)$ to cosmic strings with $\mu/\rho=4/15$ and $(-2/3,2/5)$ to domain walls with the same value of $\mu/\rho$.}
\label{fig:sdm}
\end{figure}

The solid dark matter/energy (SDE) model was first motivated by the desire to model the effects of a lattice configuration of non-Abelian cosmic strings~\cite{vil,kib}, or domain walls~\cite{bbs,fried} which would behave like an elastic solid on the scales relevant to cosmological perturbations. On dimensional grounds, the density of static cosmic strings scales as $\rho_{\rm str}\propto a^{-2}$ and that for domain walls is given by  $\rho_{\rm dw}\propto a^{-1}$. Hence, they are natural candidates for dark energy with $w=-1/3$ and $w=-2/3$ respectively. Recently, it has been shown~\cite{BCCM} that under some reasonable, but not totally general, circumstances $\mu/\rho=4/15$ for static Dirac-Nambu-Goto strings and domain walls providing sufficient rigidity to achieve stability in both cases. Hence, $(w,c_{\rm s}^2)=(-1/3,1/5)$ represents the cosmic string model and $(-2/3,2/5)$ domain walls. What is not totally obvious is that the model presented in ref.~\cite{BS}, for the which the scalar sector of perturbations is completely specified by $(w,c_{\rm s}^2)$, also naturally models a $\Lambda$ fluid with $w=-1$ and CDM with $(0,0)$. This means that the inclusion of these two parameters provides a parameterization of constant dark energy models with four interesting special cases as shown in Fig.~\ref{fig:sdm}. The aim of this paper is constrain this parameterization using a range of cosmological observations.
 
\section{Methodology}
\label{sect:meth}

An important feature of dark energy models with $w\ne -1$ is that, in order to make accurate predictions, one has to model the perturbations~\cite{CDS}. This has been studied in the context of models based on a scalar field by a number of authors~\cite{WL,BD}. In the synchronous gauge and for constant $w$, the perturbations in an elastic solid satisfy~\cite{BS,BM2}
\begin{eqnarray}
\dot\delta_{\rm s}&=&-(1+w)\left(\theta_{\rm s}+{1\over 2}\dot h\right)-3H(c_{\rm s}^2-w)\delta_{\rm s}\,,\\
\dot\theta_{\rm s}&=&-(1-3w)H\theta_{\rm s}+{w\over 1+w}k^2\delta_{\rm s}-{2w\over 3(1+w)}k^2\Pi_{\rm s}\,,
\end{eqnarray}
where $H=\dot a/ a$, $h$ is the trace of the spatial metric perturbation and the anisotropic stress is given by
\begin{equation}
\Pi_{\rm s}={3\over2w}(c_{\rm s}^2-w)\left[-\delta_{\rm s}+3(1+w)(\eta-\eta_{\rm I})\right]\,.
\end{equation}
$\eta-\eta_I$ is the change in the anisotropic part of the spatial metric component since the time when the perturbations were created, which in the case the adiabatic perturbations under consideration here is conformal time $\tau=0$. It is clear that these expression revert to those of a perfect fluid when $c_{\rm s}^2=w$ and, in particular, they model a CDM fluid $(w,c_{\rm s}^2)=(0,0)$. We will denote the energy density in the SDE relative to critical as $\Omega_{\rm s}$ and for $w=-1$ this will correspond to $\Omega_{\Lambda}$.

The equations were incorporated into {\tt CAMB}~\cite{CAMB}, which is a modified version of {\tt CMBFAST}~\cite{SZ}, and this was extensively tested. We note that the viability of topological defect models has been considered before in ref.\cite{MCMS}, where it was argued that the perturbations in SDE would behave like those in Quintessence and therefore used the perturbation model of ref.\cite{WL}. Not surprisingly, this is true near $w=-1$, but becomes a poor approximation near $w=-1/3$.

This was then incorporated into the October 2004 version of {\tt COSMOMC}~\cite{LB} in order to create Markov-Chain-Monte-Carlo (MCMC) chains which were used to estimate the confidence limits on the cosmological parameters. Up to 8 parameters were used in the fit and these were chosen to correspond to some of the known degenerate directions in the cosmological parameter space. In particular, we used $\Omega_{\rm b}h^2$, $\Omega_{\rm c}h^2$, $\tau_{\rm R}$, $\log(10^{10}A_{\rm S})$, $n_{\rm S}$, $w$, $\log c_{\rm s}$ and $\theta_{\rm A}$ the acoustic scale defined by the ratio of the sound horizon to the angular diameter distance at the redshift of recombination. The intrinsic flat priors, listed in Table \ref{flatpriors}, were chosen to be sufficient broad to incorporate the lines of degeneracy known to exist within the space of parameters and so that the maxima were far from the edges.

\begin{table}
\label{flatpriors}
\begin{center}
\begin{tabular}{|c|c|} \hline
Parameter & Prior \\ \hline
$\Omega_{\rm b} h^2 $ & (0.005,0.1) \\ 
$\Omega_{\rm c} h^2 $ & (0.01,0.99) \\ 
$\theta_{\rm A}  $ & (0.5,10) \\ 
$\tau_{\rm R} $ & (0.01,0.9) \\ 
$n_{\rm S} $ & (0.5,1.5) \\ 
$\log (10^{10} A_{\rm S})$ & (2.7,5.0) \\ 
$w $ & (-1.33, 0.33) \\ 
$\log c_{\rm s} $ & (-5,0) \\ \hline
\end{tabular}
\end{center}
\caption{Table of flat priors. The notation $(a,b)$ for a particular parameter gives the lower and upper bounds allowed in the fit. Note that the parameters $w$ and $c_{{\rm s}}^{2}$ are also bounded by the constraint that $0 \le  c_{{\rm v}}^{2} \le 1$. $w=-4/3$ is the lowest value of $w$ allowed by this constraint.}
\end{table}

We considered six different models, denoted I, II, III, IV, V and VI in the subsequent text. I is the standard $\Lambda$CDM model with $w=-1$; II is the full eight parameter fit to the data allowing both $w$ and $\log c_{\rm s}$ to vary; III has $w=-1/3$ (fixed) corresponding to string model with $\log c_{\rm s}$ allowed to vary; IV is the same as III but with $w=-2/3$, corresponding to domain walls; the final two models, V and VI, use fixed values of $(w,c_{\rm s}^2)$, $(-1/3,1/5)$ and $(-2/3,2/5)$ respectively, which are those predicted for static Dirac-Nambu-Goto strings and domain walls with $\mu/\rho=4/15$~\cite{BCCM}. Models I, V, and VI have six parameters, III and IV have seven, and II has eight.

The following procedure was used to ensure convergence of the MCMC chains: first, a single chain was run for a short period (less than a day) and covariance matrix was computed from it. This was used as input for a further five chains which were run until the variance between the chains of the largest normalized eigenvalue of the covariance matrix was less than 0.1, a process which took a further one to two days. On a small number of occasions, it was necessary to run a further five chains using the covariance matrix of the earlier five chains. We typically obtained a variance in the normalized eigenvalue which was between 0.01 and 0.05 for each model considered. Such a low variance ensures that each of the five chains agree on the degenerate directions within the parameter space.

We will use three different kinds of data in our fits. Cosmic Microwave Background (CMB) data from the 1st year observations of the Wilkinson Microwave Anisotropy Probe (WMAP)~\cite{Bennett} (likelihood computed using the routine provided by the WMAP team~\cite{verde}) and the three experiments which observe at substantially higher resolution than possible using WMAP. These are the Very Small Array (VSA)~\cite{VSA}, the Cosmic Background Imager (CBI)~\cite{CBI} and the ArCminute Bolometer ARray (ACBAR)~\cite{acbar}. The other two main datasets which were used were Large-Scale Structure (LSS) information from the 2DF Galaxy Redshift Survey (2DFGRS), using data in the redshift range $0.02 < k/(h{\rm Mpc^{-1}}) < 0.15$ where non-linear effects are found to be negligible~\cite{perc}, and measurements of the brightness of type Ia Supernovae (SNe). For the SNe analysis, the set of 42 SNe were used from the Supernovae Cosmology Project~\cite{perm}, at redshifts between 0.18 and 0.23, and 18 from the Cal$\acute{\rm a}$n/Toledo Supernoave Survey~\cite{hamuy}, at redshifts below 0.1. Using these a sample of 54 were selected which form the primary (fit C) dataset as described in~\cite{perm}. Analytic marginalization was then performed over the absolute magnitude by comparing the effective apparent magnitude with luminosity distance evaluated at the relevant survey redshift. In section \ref{sect:disc} we also investigate the implications of  the Hubble Space Telescope (HST) key project measurement of $h=0.72\pm 0.08$~\cite{hst} and that of $\Omega_{\rm b}h^2=0.020\pm 0.002$ which comes from considerations of Big Bang Nucleosynthesis (BBN)~\cite{bbn}.

The interpretation  of the LSS data in these SDE models needs to be modified. We have already pointed out that one of the defining properties of dark energy is a large Jean's length, $\lambda_{\rm J}$, at the present epoch. In these models, $\lambda_{\rm J}\sim c_{\rm s}\tau_0$, where $\tau_0$ is the conformal time today and therefore if $c_{\rm s}$ is small (or zero) we can expect the SDE to cluster. It is clear that it would be sensible to avoid it being clustered on scales $<10{\rm Mpc}$, which implies $c_{\rm s}>0.001$, since it would contribute to dynamical estimates of $\Omega_{\rm m}$ from rich clusters of galaxies which are sub-critical (see, for example, ref.~\cite{dyn}).

However, in principle there is nothing to prevent it clustering on larger scales and it is desirable to constrain this possibility using the data. The 2DFGRS measures the power spectrum of the galaxies which is usually assumed to be related to that of the underlying density field via a linear bias factor, commonly believed to be close to one, via the relation $P_{\rm g}(k)=b^2P_{\rm c}(k)$ where $P(k)$ is the power spectrum of the relevant component. If there is an additional SDE component in the universe which is clustered in some way, then it will also effect the power spectrum of the galaxies. In order to attempt to take this into account,  we will use 
\begin{equation}
P_{\rm g}(k)=b^2\bigg(\Omega_{\rm s}P_{\rm s}(k)+\Omega_{\rm c}P_{\rm c}(k)\bigg)\,,
\end{equation}
which makes the, untested, assumption that the biasing of the SDE and CDM is in proportion to their density. This will play an important role in our results since CMB observations prefer low values of $c_{\rm s}$, but these are clearly ruled out by 2DFGRS.

The number of WMAP data points fitted to are 1348 from (899 from temperature and 449 from the temperature polarization cross-correlation), 16 from VSA, 13 from CBI, 7 from ACBAR, 32 from 2DFGRS and 54 from SNe. Although we will present the number of datapoints, $N$, involved in our fits in order to attempt to assess the necessity or otherwise of the parameters which we have added, some caution should be attached to them since in a number of cases the errorbars are correlated and these correlations need to be taken into account. The number of degrees of freedom computed in this way is a upper bound.

\begin{table}[]
\label{tab:cmb}
\begin{center}
\begin{tabular}{|c|c|c|c|c|c|c|} \hline
& \multicolumn{6}{|c|}{Model} \\ \hline 
 Parameter &  I &  II &  III &  IV &  V &  VI \\ \hline

$\Omega_{\rm b} h^2 $ & $0.023 \pm 0.002$ & $0.028 \pm 0.003$ & $0.028 \pm 0.003$ & $0.026 \pm 0.003$ & $0.024 \pm 0.002$ & $0.024 \pm 0.002$ \\ \hline

$\Omega_{\rm c} h^2 $ & $0.103 \pm 0.014 $ & $0.067 \pm 0.015$ & $0.066 \pm 0.014$ & $0.079 \pm 0.016$ & $0.097 \pm 0.013$ & $0.093 \pm 0.016$ \\ \hline

$\theta_{\rm A}  $ & $1.043 \pm 0.006 $ & $1.050 \pm 0.006$ & $1.050 \pm 0.006$ & $1.048\pm 0.007$ & $1.044 \pm 0.006$ & $1.044 \pm 0.006$ \\ \hline

$\tau_{\rm R} $ & $0.17 \pm 0.11 $ & $0.46 \pm 0.11$ & $0.47 \pm 0.10 $ & $0.37 \pm 0.13$ & $0.16 \pm 0.08$ & $0.23 \pm 0.12$ \\ \hline

$n_{\rm S} $ & $0.98 \pm 0.05$ & $1.13 \pm 0.07$  & $1.13 \pm 0.06$ & $1.08 \pm 0.07$ & $1.00 \pm 0.05$ & $1.02 \pm 0.07$ \\ \hline

$\log (10^{10} A_{\rm S})$ & $3.2 \pm 0.2$ & $3.7 \pm 0.2$ & $3.7 \pm 0.2$ & $3.5 \pm 0.2$ & $3.1 \pm 0.2$ & $3.3 \pm 0.2$ \\ \hline

$w $ & -1 & $-0.38 \pm 0.16 $ & -1/3 & -2/3 & -1/3 & -2/3 \\ \hline

$\log c_{\rm s} $ & - & $<-1.0$ & $<-1.1$ & $<-0.7$ & -0.35  & -0.2 \\ \hline \hline

$\Omega_{\rm m} $ & $0.23 \pm 0.06$ & $0.30 \pm 0.10$ & $0.31 \pm 0.06$ & $0.22 \pm 0.07$ & $0.46 \pm 0.06$ & $0.28 \pm 0.08$ \\ \hline 

$\Omega_{\rm s} $ & $0.77 \pm 0.06$ & $0.70 \pm 0.10$ & $0.69 \pm 0.06$  & $0.78 \pm 0.07$ & $0.54 \pm 0.06$ & $0.72 \pm 0.08$ \\ \hline

$t_{0}/{\rm Gyr}  $ & $13.5 \pm 0.4$ & $14.1 \pm 0.7$ & $14.2 \pm 0.3$ & $13.4 \pm 0.4$ & $14.5 \pm 0.3$ & $13.7 \pm 0.4$ \\ \hline 

$z_{\rm re}  $ & $16 \pm 6$ & $28 \pm 3$ & $28 \pm 3$ & $25 \pm 4$ & $16 \pm 5$ & $19 \pm 6$ \\ \hline 

$h  $ & $0.76 \pm 0.08$ & $0.58 \pm 0.09$ & $0.55 \pm 0.02$ & $0.72 \pm 0.06$ & $0.52 \pm 0.01$ & $0.66\pm 0.06$ \\ \hline \hline

$-2 \log \mathcal{L}  $ & 1453 & 1443 & 1443 & 1446 & 1453 & 1451 \\ \hline \hline 

AIC & 1465 & 1459 & 1457 & 1460 & 1465 & 1463 \\ \hline
BIC & 1496 & 1501 & 1494 & 1497 & 1496 & 1494 \\ \hline \hline

\end{tabular}
\end{center}
\caption{Marginalised parameter constraints for the models I-VI using the CMB data only ($N=1384$). Note the improvement in the value of $-2\log {\cal L}$ in going from model I to model II and that the values for V and VI have similar values to I. For sure these models cannot be ruled out! The values given for $\log c_{\rm s}$ are the upper $95\%$ confidence levels in models II, III and IV. AIC and BIC refer to the Akaike and Bayesian Inference Criteria, respectively, which are discussed in section \ref{sect:disc}}
\end{table}

\begin{table}[]
\label{tab:cmb2df}
\begin{center}
\begin{tabular}{|c|c|c|c|c|c|c|} \hline
& \multicolumn{6}{|c|}{Model} \\ \hline 
 Parameter &  I &  II &  III &  IV &  V &  VI \\ \hline 

$\Omega_{\rm b} h^2 $ & $0.023 \pm 0.001$  & $0.026 \pm 0.003$ & $0.025 \pm 0.002$ & $0.025 \pm 0.002$ & $0.024 \pm 0.002$ & $0.025 \pm 0.002$ \\ \hline

$\Omega_{\rm c} h^2 $ & $0.115 \pm 0.007$ & $0.091 \pm 0.014$ & $0.091 \pm 0.013$ & $ 0.095 \pm 0.009$ & $0.097 \pm 0.011$ & $0.096 \pm 0.009$ \\ \hline

$\theta_{\rm A}  $ & $1.042 \pm 0.005$ & $1.049 \pm 0.007$ & $1.048 \pm 0.007$ & $1.047 \pm 0.007$ & $1.045 \pm 0.006$ & $1.046 \pm 0.006$ \\ \hline

$\tau_{\rm R} $ & $0.11 \pm 0.06$ & $0.29 \pm 0.12$ & $0.27 \pm 0.11$ & $0.27 \pm 0.10$ & $0.16 \pm 0.08$ & $0.23 \pm 0.10$ \\ \hline

$n_{\rm S} $ & $0.96 \pm 0.02$ & $1.05 \pm 0.08$ & $1.05 \pm 0.07$ & $1.04 \pm 0.06$ & $1.00 \pm 0.04$ & $1.02 \pm 0.06$ \\ \hline

$\log (10^{10} A_{\rm S})$ & $3.1 \pm 0.1$ & $3.4 \pm 0.2$ & $3.4 \pm 0.2$ & $3.4 \pm 0.2$ & $3.1 \pm 0.2$ & $3.3 \pm 0.2$ \\ \hline

$w $ & -1 & $-0.62 \pm 0.15$ & -1/3 & -2/3 & -1/3 & -2/3 \\ \hline

$\log c_{\rm s} $ & - & $-0.77 \pm 0.28$ & $-0.94 \pm 0.24$ & $-0.76 \pm 0.27$ & -0.35  & -0.2 \\ \hline \hline

$\Omega_{\rm m} $ & $0.28 \pm 0.03$ & $0.28 \pm 0.06$ & $0.41 \pm 0.07$ & $0.28 \pm 0.05$ & $0.46 \pm 0.05$ & $0.28 \pm 0.05$ \\ \hline 

$\Omega_{\rm s} $  & $0.72 \pm 0.03$  & $0.72 \pm 0.06$ & $0.58 \pm 0.07$ & $0.72 \pm 0.05$ & $0.54 \pm 0.05$ & $0.72 \pm 0.05$ \\ \hline

$t_{0}/{\rm Gyr} $ & $13.7 \pm 0.2$ & $13.6 \pm 0.4$ & $14.3 \pm 0.3$ & $13.6 \pm 0.4$ & $14.5 \pm 0.2$ & $13.7 \pm 0.4$ \\ \hline 

$z_{\rm re}  $ & $13 \pm 4$ & $22 \pm 5$ & $21 \pm 5$ & $21 \pm 5$ & $16 \pm 5$ & $ 19 \pm 5$ \\ \hline 

$h  $ & $0.71 \pm 0.03$ & $0.65 \pm 0.05$ & $0.53 \pm 0.02$ & $0.66 \pm 0.04$ & $0.52 \pm 0.01$ & $ 0.66 \pm 0.04 $ \\ \hline \hline

$-2 \log \mathcal{L}  $ & 1488 & 1477 & 1484 & 1479 & 1490 & 1482 \\ \hline \hline

AIC & 1500 & 1493 & 1495 & 1493 & 1502 & 1494 \\ \hline
BIC & 1532 & 1537 & 1535 & 1530 & 1534 & 1526 \\ \hline \hline

\end{tabular}
\end{center}
\caption{As for table \ref{tab:cmb} but using the CMB and 2DFGRS datasets ($N=1416$). Now the value of $\log c_{\rm s}$ is constrained and is not just an upper bound.}
\end{table}

\begin{table}[]
\label{tab:cmbsn}
\begin{center}
\begin{tabular}{|c|c|c|c|c|c|c|} \hline
& \multicolumn{6}{|c|}{Model} \\ \hline 
 Parameter &  I &  II &  III &  IV &  V &  VI \\ \hline 

$\Omega_{\rm b} h^2 $ & $0.023 \pm 0.001$ & $0.027 \pm 0.003$ & $0.029 \pm 0.002$ & $0.027 \pm 0.002$ & $0.025 \pm 0.002$ & $0.026 \pm 0.003$ \\ \hline

$\Omega_{\rm c} h^2 $ & $0.108 \pm 0.011$ & $0.073 \pm 0.017$ & $0.056 \pm 0.010$ & $0.071 \pm 0.012$ & $0.086 \pm 0.014$ & $0.080 \pm 0.015$ \\ \hline

$\theta_{\rm A}  $ & $1.042 \pm 0.005$ & $1.049 \pm 0.007$ & $1.052 \pm 0.006$ & $1.049 \pm 0.006$ & $1.046 \pm 0.006$ & $1.047 \pm 0.007$ \\ \hline

$\tau_{\rm R} $ & $0.13 \pm 0.07$ & $0.42 \pm 0.13$ & $0.54 \pm 0.07$ & $0.43 \pm 0.09$ & $0.22 \pm 0.09$ & $0.31 \pm 0.12$ \\ \hline

$n_{\rm S} $ & $0.96 \pm 0.04$ & $1.11 \pm 0.08$ & $1.17 \pm 0.05$ & $1.11 \pm 0.06$ & $1.04 \pm 0.06$ & $1.06 \pm 0.07$ \\ \hline

$\log (10^{10} A_{\rm S})$ & $3.1 \pm 0.1$ & $3.6 \pm 0.2$ & $3.8 \pm 0.1$ & $3.6 \pm 0.2$ & $3.2 \pm 0.2$ & $3.4 \pm 0.2$ \\ \hline

$w $ & -1 & $-0.68 \pm 0.14$ & -1/3 & -2/3 & -1/3 & -2/3 \\ \hline

$\log c_{\rm s} $ & - & $<-0.7$ & $<-1.2$ & $<-0.7$ & -0.35  & -0.2 \\ \hline \hline

$\Omega_{\rm m} $ & $0.25 \pm 0.05$ & $0.19 \pm 0.05$ & $0.27 \pm 0.04$ & $0.18 \pm 0.05$ & $0.40 \pm 0.07$ & $0.22 \pm 0.06$ \\ \hline 

$\Omega_{\rm s} $ & $0.75 \pm 0.05$ & $0.81 \pm 0.05$ & $0.73 \pm 0.04$ & $0.82 \pm 0.05$ & $0.60 \pm 0.07$ & $0.78 \pm 0.06$ \\ \hline

$t_{0}/{\rm Gyr}  $ & $13.6 \pm 0.3$ & $13.3 \pm 0.4$ & $14.1 \pm 0.3$ & $13.2 \pm 0.4$ & $14.5 \pm 0.3$ & $13.5 \pm 0.4$ \\ \hline 

$z_{\rm re}  $ & $14 \pm 5$ & $26 \pm 4$ & $29 \pm 1$ & $26 \pm 3$ & $18 \pm 5$ & $22 \pm 5$ \\ \hline 

$h  $ & $0.73 \pm 0.05$ & $0.74 \pm 0.08$ & $0.57 \pm 0.02$ & $0.75 \pm 0.05$ & $0.53 \pm 0.02$ & $ 0.71 \pm 0.06$ \\ \hline \hline

$-2 \log \mathcal{L}  $ & 1506 & 1500 & 1509 & 1500 & 1524 & 1508 \\ \hline \hline

AIC & 1518 & 1516 & 1523 & 1514 & 1536 & 1520 \\ \hline
BIC & 1550 & 1558 & 1560 & 1551 & 1568 & 1552 \\ \hline \hline

\end{tabular}
\end{center}
\caption{As for table \ref{tab:cmb} but using CMB and SNe data ($N=1438$). The improvement in the fit in going from I to II is less marked when compared to tables II and III, but is still reasonably significant. Model V is ruled out in this case, but VI is not.}
\end{table}

\begin{table}[!htb]
\label{tab:all}
\begin{center}
\begin{tabular}{|c|c|c|c|c|c|c|} \hline
& \multicolumn{6}{|c|}{Model} \\ \hline 
 Parameter &  I &  II &  III &  IV &  V &  VI \\ \hline 

$\Omega_{\rm b} h^2 $ & $0.023 \pm 0.001$ & $0.024 \pm 0.002$ & $0.027 \pm 0.002$ & $0.025 \pm 0.002$ & $0.024 \pm 0.002$ &  $0.025 \pm 0.002$ \\ \hline

$\Omega_{\rm c} h^2 $ & $0.116 \pm 0.006$ & $0.102 \pm 0.014$  & $0.082 \pm 0.11$ & $0.092 \pm 0.008$ & $0.090 \pm 0.010$ & $0.093 \pm 0.008$ \\ \hline

$\theta  $ & $1.043 \pm 0.005$ & $1.045 \pm 0.006$ & $1.051 \pm 0.007$ & $1.048 \pm 0.006$ & $1.046 \pm 0.006$ & $1.048 \pm 0.006$ \\ \hline

$\tau_{\rm R} $ & $0.11 \pm 0.05$ & $0.21 \pm 0.12$ & $0.33 \pm 0.10$ & $0.29 \pm 0.10$ & $0.19 \pm 0.09$ & $0.27 \pm 0.09$ \\ \hline

$n_{\rm S} $ & $0.96 \pm 0.02$ & $1.01 \pm 0.06$ & $1.09 \pm 0.06$ & $1.05 \pm 0.06$ & $1.02 \pm 0.05$ & $1.04 \pm 0.06$  \\ \hline

$\log (10^{10} A_{\rm S})$ & $3.1 \pm 0.1$ & $3.3 \pm 0.2$ & $3.5 \pm 0.2$ & $3.4 \pm 0.2$ & $3.2 \pm 0.2$ & $3.4 \pm 0.2$ \\ \hline

$w $ & -1 & $-0.80 \pm 0.15$ & -1/3 & -2/3 & -1/3 & -2/3 \\ \hline

$\log c_{\rm s} $ & -  & $-0.87 \pm 0.42$  & $-0.91 \pm 0.19$  & $-0.71 \pm 0.26$ & -0.35  & -0.2 \\ \hline \hline

$\Omega_{\rm m} $ & $0.28 \pm 0.03$ & $0.27 \pm 0.04$ & $0.37 \pm 0.06$ & $0.26 \pm 0.04$ & $0.42 \pm 0.05$ & $0.26 \pm 0.04$ \\ \hline 

$\Omega_{\rm s} $ & $0.72 \pm 0.03$ & $0.73 \pm 0.04$ & $0.63 \pm 0.06$ & $0.74 \pm 0.04$ & $0.58 \pm 0.05$ & $0.74 \pm 0.04$ \\ \hline

$t_{0}/{\rm Gyr}  $ & $13.7 \pm 0.2$ & $13.6 \pm 0.3$ & $14.2 \pm 0.3$ & $13.5 \pm 0.4$ & $14.5 \pm 0.3$ & $13.5 \pm 0.3$ \\ \hline 

$z_{\rm re}  $ & $13 \pm 4$ & $18 \pm 6$ & $24 \pm 4$ & $22 \pm 4$ & $17 \pm 5$ & $21 \pm 4$ \\ \hline 

$h  $ & $0.70 \pm 0.03$ & $0.68 \pm 0.04$ & $0.54 \pm 0.02$ & $0.68 \pm 0.04$ & $0.52 \pm 0.02$ & $ 0.67 \pm 0.04$ \\ \hline \hline

$-2 \log \mathcal{L}  $ & 1541 & 1535 & 1552 & 1535 & 1561 & 1538 \\ \hline \hline

AIC & 1553 & 1550 & 1566 & 1548 & 1573 & 1550 \\ \hline
BIC & 1585 & 1593 & 1603 & 1586 & 1605 & 1582 \\ \hline \hline

\end{tabular}
\end{center}
\caption{As for table \ref{tab:cmb}, but using CMB, 2DFGRS and SNe data ($N=1470)$.}
\end{table}

\section{Results}
\label{sect:res}

We will present constraints on the various parameters in the models I-VI for four different combinations of datasets. Firstly, we will just use the CMB dataset (WMAP+VSA+CBI+ACBAR) alone, then we will include either 2DFGRS or SNe and finally we consider the combination of all three. The detailed results of these different options are presented in tables II, III, IV and V. The results of model I can be seen to be compatible with earlier estimates (for example, ref.~\cite{spergel}) and calibrate the results of model II and models III-VI which are restrictions of it to various special cases.

\begin{figure}
\centering
\mbox{\resizebox{0.7\textwidth}{!}{\includegraphics{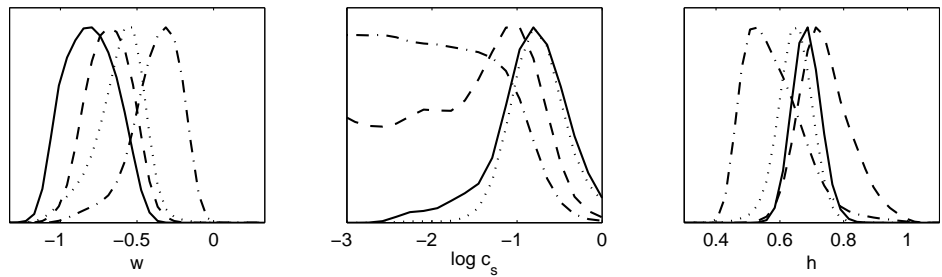}}}
\caption{1D likelihood plots for $w$, $\log c_{\rm s}$ and $h$. The curves are for the four different combinations of data discussed in the text: CMB (dash-dot line), CMB + 2DFGRS (dotted line) and CMB + SNe (dashed line) and CMB + 2dF + SNe (solid line). Notice that the preferred value of $w$ moves closer to -1 as more datasets are included. Also that low values of $c_{\rm s}$ are excluded by the 2DFGRS data.}
\label{fig:1dplots}
\end{figure}

1D likelihood curves for $w$, $\log c_{\rm s}$ and  $h$ using model II are presented in Fig.~\ref{fig:1dplots} and these illustrate the main results of our analysis. We see that using the CMB data alone the best fitting values of $w=-0.38\pm 0.16$ and that low values of $c_{\rm s} (<0.1$ at 95\% confidence) are preferred. There is an improvement in $\chi^2=-2\log{\cal L}$ of $\sim 10$ from model I with the inclusion of two extra parameters. This results in a change in the best-fitting value of $h=0.58\pm 0.09$ which is compatible with the HST key project value at the 1-$\sigma$ level, but is lower than usually assumed.

There is a reason for these particular values being favoured. As pointed out in ref.~\cite{bbs} the only way the SDE, Quintessence and $\Lambda$ models can be distinguished using CMB data alone is via the Integrated-Sachs-Wolfe (ISW) contribution to the anisotropy, which is created by the decay of gravitational potentials along the line of sight. Since the models react to metric perturbations in a different ways, the ISW contribution to the anisotropy is necessarily model-dependent. It was shown that SDE models with $w=-1/3$ and $c_{\rm s}=0$ have anisotropy spectra which turn down on large angular scales; a known feature of the WMAP (and COBE) datasets. It is for this reason that this particular choice of parameters is preferred in this case. This apparent anomaly may be  an artifact of, for example, imperfect foreground subtraction, but taking the WMAP data at face value one is lead to the  conclusion that this model is the best-fit. We note that this is somewhat contrary to the standard dogma since the anisotropy spectrum in the $\Lambda$CDM model increases on the largest scales.

The inclusion of the 2DFGRS dataset into the fit shifts the preferred value to $w=-0.62\pm 0.15$ and dramatically constrains $\log c_{\rm s}=-0.77 \pm 0.28$. The value of $h=0.65\pm 0.05$ and $\chi^2$ is still $\sim 10$ smaller than in the $w=-1$ fixed case. The reason for these changes in the preferred values is due to two effects. The first is that the 2DFGRS survey measures the cosmological parameter combination $\Omega_{\rm m}h \sim 0.2$ modifying the preferred values of $\Omega_{\rm m}$ and $h$ with the consequent effect on $w$ via the relation for $\theta_{\rm A}$, and also the fact that if $c_{\rm s}$ is low then the SDE contributes to $P_{\rm g}(k)$ in the relevent range of $k$. A trade-off between this effect, and the desire to keep $c_{\rm s}$ as low as possible in order to improve the fit to the large angular scale CMB, results in the preferred value of $c_{\rm s}$.

The SNe dataset is known to prefer values of $w$ which are closer to -1 and this is borne out by its inclusion in the fit, both in the case when it is only considered in conjunction with the CMB data and also with 2DFGRS. We find that $w=-0.68\pm 0.14$ for CMB + SNe and $w=-0.80\pm 0.15$ for CMB + 2DFGRS + SNe. In these cases the improvement in the value of $\chi^2$, when compared to $w=-1$ fixed,  is less, $\sim 7$, than for CMB data alone and CMB + 2DFGRS.

\begin{figure} 
\centering
\mbox{\resizebox{0.8\textwidth}{!}{\includegraphics{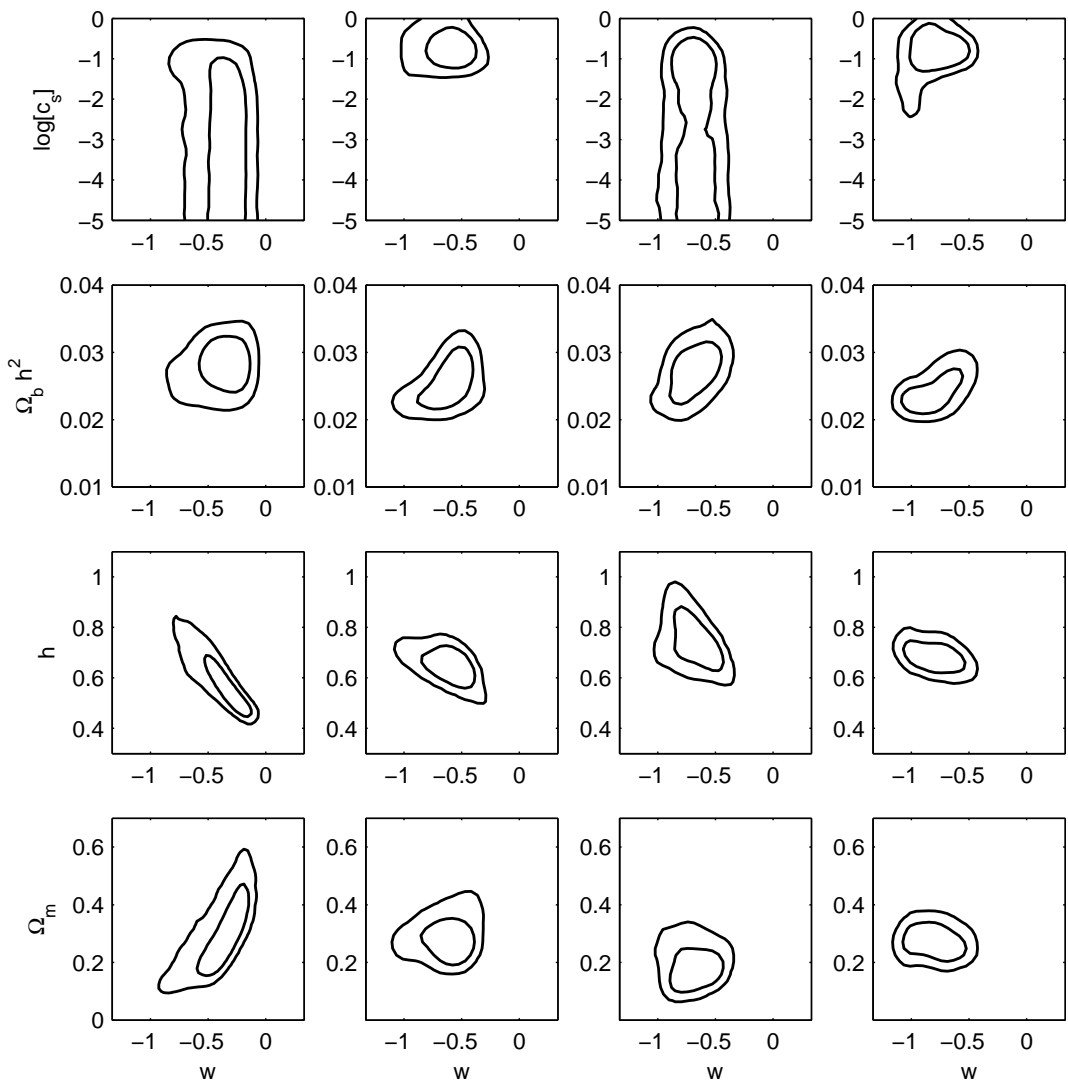}}}
\caption{A selection 2D likelihood contours in model II. The four different columns correspond to the different data combinations discussed in the text: CMB (left column), CMB + 2DFGRS (middle-left column) and CMB + SNe (middle-right column) and CMB + 2DFGRS + SNe (right column).}
\label{fig:2dplots}
\end{figure}

A selection of 2D likelihoods are presented in Fig.~\ref{fig:2dplots} which illustrate the degeneracies between $w$ and the other parameters. These bear out our earlier comments. We see that the preferred value of $\log c_{\rm s}$ is independent of $w$ , but that the preferred values of $\Omega_{\rm m}$ and $h$ are degenerate with $w$ using CMB data alone. This degeneracy is broken by the inclusion of data from the 2DFGRS and SNe, so much so that when both are included, the preferred value of $w$ appears to be independent of them. We also see that $w$ appears not to be degenerate with $\Omega_{\rm b}h^2$ for the CMB data alone, but when all the datasets are used together a degeneracy appears. We will return to this issue in section \ref{sect:disc}.

The models III and IV model dark energy fluids which scale like static cosmic strings $(w=-1/3)$ and domain walls $(w=-2/3)$ respectively without restricting to  values of $c_{\rm s}$ corresponding to $\mu/\rho=4/15$. We see that for CMB data alone that the best-fitting parameters for model III are very close to those of model II, and that $\chi^2\sim 10$  better than model I for the inclusion of a single parameter. This parameter, $\log c_{\rm s}$, allows a better fit to the data by reducing the power at large scales as discussed earlier. Model IV is also a better fit to the data than model I. When the 2DFGRS data and SNe are included, by themselves and together, model IV becomes a better fit to the data in keeping with the results found for model II.

We have also explored models V and VI which have $w$ fixed and $\mu/\rho=4/15$. These models have the same number of parameters as model I and therefore the values of $\chi^2$ can be compared directly. We see that in the case of CMB data alone that models I and V fit equally as well, whereas model VI provides a marginally better fit. When 2DFGRS is also included the fit of model VI is significantly improved over model I; model V is a less good fit, but only by a small amount. SNe data rejects $w=-1/3$ and this is borne out by our results. Model V provides a significantly poorer fit to the data when SNe is included, either by itself or in conjunction with 2DFGRS, whereas model VI is a less good fit than model I for CMB + SNe, but is a better fit for CMB + 2DFGRS + SNe.

\section{Discussion and Conclusions} 
\label{sect:disc}

Our analysis of the SDE parameter space has turned up some interesting features. We have found that the cosmic string model with $w=-1/3$ and $\mu/\rho=4/15$  is only excluded when SNe data is included. If $c_{\rm s}$ is allowed to vary then the poorer fit to the SNe data can be offset by a much better fit to the large angular scale CMB data. It appears that the $\Lambda$ based model and the domain wall model appear to fit the data equally well at this stage, dependent on ones choice of data. Given the dependence of our results on the choice of data and the possible systematic effects which may be hidden within the datasets we feel that the main conclusion is that, of the four models discussed in section \ref{intro}, only one (the pure CDM model $(w,c_{\rm s}^2)=(0,0)$) can be ruled out with any great certainty, and that the models based on topological defects are compatible with the present data. The domain wall model is probably the most favoured, but not with considerable significance.

The current established approaches to the cosmological parameters (CMB, LSS and SNe) are unlikely to separate them in the near future. The resolving power of SNe data may be improved by finding more SNe and covering a larger redshift range (see, for example, refs.~\cite{WA1,WA2}), but the CMB angular power spectra and linear LSS power spectrum at low redshift have limited room for improvement in constraining $w$~\cite{bbs,Huey}. WMAP polarization data may improve the situation by reducing the uncertainty on $\tau_{\rm R}$, but the resolving power of total intensity observations has largely been exhausted. It is clear that radical alternatives to the dark energy problem are required and many new cosmological tests using upcoming data have been proposed. These include: galaxy cluster surveys traced by the Sunyaev-Zel'dovich effect and X-ray emission~\cite{HHM,BW2}; baryonic oscillations in the matter power spectrum~\cite{HE}; cross-correlation of the CMB with measures of LSS~\cite{CT}; and weak lensing~\cite{wkl1}. In some cases these are already beginning to have an impact (see, for example, refs.~\cite{ein,BC}).

We should comment on the reasons why some earlier analyses might appear to give constraints on $w$ which are somewhat closer to -1 than those we present here. Some of this is due to the fact that perturbations in the SDE model evolve differently to scalar field motivated models, although this effect is only significant away from $w=-1$. The main reason is use of priors on $h$ and $\Omega_{\rm b}h^2$ from HST and BBN which may contain systematic errors. This is apparent from carefully examining the 2D likelihoods presented in Fig.~\ref{fig:2dplots} or noting that the best-fitting values of $h$ and $\Omega_{\rm b}h^2$ presented in tables II, III, IV and V clearly depend on the specific model. In order to reinforce this we have performed importance sampling on our MCMC chains. This was done by using the covariance matrix from the earlier well-converged chains (normalized variance in the eigenvalues typically between 0.01 and 0.05) to run further chains generating approximately ten to twenty thousand samples from which to importance sample. 

The best-fitting values of $w$ are presented in table \ref{tab:is} for a range of constraints on $h$ to test the sensitivity to possible systematic errors in the measurement of the Hubble constant, with and without a prior on $\Omega_{\rm b}h^2$. We see that increasing the central value of the external constraint on $h$ reduces the value of $w$ toward -1. This is a strong effect when one consider the CMB data alone, but this degenerate direction is already partially constrained when the other datasets are included. The addition of the constraint from BBN has an effect in the same direction. If one uses data from CMB+2DFGRS+SNe+HST+BBN then one comes to the conclusion that $w=-0.89\pm 0.11$, but the possibility of their being no systematic effects in these five datasets seems a little unlikely. We believe that constraints from two, possible three different datasets, are much more plausible and in just about every possible combination of the datasets, models with $w$ significantly different from -1 are favoured.

\begin{table}
\label{tab:is}
\begin{center}
\begin{tabular}{|c|c|c|c|c|c|c|c|c|} \hline
& \multicolumn{2}{|c|}{CMB} & \multicolumn{2}{|c|}{CMB+2DFGRS}  & \multicolumn{2}{|c|}{CMB+SNe}  & \multicolumn{2}{|c|}{CMB+2DFGRS+SNe} \\ \hline 

 $h$ & NO BBN & BBN & NO BBN & BBN & NO BBN & BBN & NO BBN & BBN \\ \hline
$0.56 \pm 0.08$ & $-0.35 \pm 0.12$ & $-0.47 \pm 0.18$ & $-0.58 \pm 0.14$ & $-0.67 \pm 0.16$ & $-0.60 \pm 0.14$ & $-0.76 \pm 0.12$ &  $-0.76 \pm 0.15$ &  $-0.83 \pm 0.12$ \\ \hline 

$0.64 \pm 0.08$ & $-0.43 \pm 0.14$ & $-0.61 \pm 0.20$ &  $-0.63 \pm 0.16$&  $-0.76 \pm 0.17$ &  $-0.64 \pm 0.14$ & $-0.79 \pm 0.12$&  $-0.78 \pm 0.15$ &  $-0.86 \pm 0.12$ \\ \hline 

$0.72 \pm 0.08$ & $-0.53 \pm 0.15$ & $-0.74 \pm 0.18$ & $-0.69 \pm 0.18$ & $-0.85 \pm 0.17$ & $-0.67 \pm 0.13$ & $-0.82 \pm 0.11$ & $-0.80 \pm 0.15$ & $-0.89 \pm 0.11$ \\ \hline 

$0.80 \pm 0.08$ & $-0.63 \pm 0.16$ & $-0.84 \pm 0.15$ & $-0.75 \pm 0.20$ & $-0.93 \pm 0.15$ & $-0.71 \pm 0.12$ & $-0.85 \pm 0.10$ & $-0.82 \pm 0.15$ & $-0.92 \pm 0.11$ \\ \hline 

\end{tabular}
\end{center}
\caption{Constraints on $w$ due external constraints on $h$ and $\Omega_{\rm b}h^2$. We have used different constraints on $h$ to represent possible systematic effects in its measurement. The HST key project value is $h=0.72\pm 0.08$ and BBN implies that $\Omega_{\rm b}h^2=0.020\pm 0.002$.}
\end{table}

We have also assessed whether the inclusion of extra parameters in models II, III and IV are justified by the Akaike~\cite{aic} and Bayesian Inference Criteria~\cite{bic}, AIC and BIC respectively, as suggested in ref.~\cite{liddle}. The definitions are ${\rm AIC}=-2\log{\cal L}+2k$ and ${\rm BIC}=-2\log{\cal L}+k\log N$, where $k$ is the number of parameters in the model. If one computes these statistics for models with different numbers of parameters, constrained by the same data, then the one with the lowest value of the statistic is preferred. This can be used to justify the necessity for the extra parameters included in the fit, although the fact that there are two different statistics prevents one coming to totally unequivocal conclusions! It is clear that the improvement in $\chi^2$ for a BIC is much larger since it requires an improvement of $\sim 7$ for each parameter added, whereas the AIC only requires an improvement of $2$. The results presented in tables II, III, IV and V are somewhat inconclusive with AIC and BIC leading to somewhat contradictory conclusions. For example, using the CMB + 2DFGRS data model II does considerably better than model I with respect to AIC, but the opposite is true for BIC. However, there is some evidence that model IV which has one extra parameter would be selected over model I by both AIC and BIC. Clearly more specialized tests would be required to come to an unequivocal conclusion on this issue.
 
An issue which we have not addressed in this paper is the possibility of intrinsic anisotropy due to the structure of  SDE. Our equations of motion for the SDE implicitly assume that the distribution of the SDE is initially isotropic, and hence the elasticity tensor is specified by two moduli, the bulk modulus (which is related to $w$) and the rigidity modulus $\mu$. It is likely that SDE models based on cosmic strings and domain walls will have point symmetry which will require additional moduli. It is possible that these could lead to a reduction in the large angular scale CMB anisotropies and the intrinsic alignment believed to exist between different multipoles~\cite{teg}. Consideration of these effects could further improve the fits of these models to the current data and we are currently exploring this.

\section*{Acknowledgments}

We would like to thank Brandon Carter and Elie Chachoua for their collaboration on earlier work which motivated this study. The numerical calculations were performed using COBRA ({\tt http://www.jb.man.ac.uk/research/cobra}) at Jodrell Bank Observatory. We thank the staff and the other users of this facility for their assistance. We would like to thank Anthony Lewis for making {\tt CAMB} and {\tt COSMOMC} publically available.


\begin{thebibliography}{99}
 
\newcommand{\prlet}{Phys.\ Rev.\ Lett.}
\newcommand{\npb}{Nucl.\ Phys.\ B}
\newcommand{\pletb}{Phys.\ Lett.\ B}
\newcommand{\prevd}{Phys.\ Rev.}
\newcommand{\jhep}{J.\ High Energy Phys.}
\newcommand{\cqg}{Class.\ Quant.\ Grav.}
\newcommand{\jast}{Ap. J.}

\bibitem{spergel}
D.N. Spergel et al, {\em Ap. J. Suppl.} {\bf 148}, 175 (2003)

\bibitem{tegmark}
M. Tegmark et al, {\em \prevd} {\bf D69}, 103501 (2004)

\bibitem{PR}
B. Ratra and P.J.E. Peebles, {\em \prevd} {\bf D37}, 3406 (1998)

\bibitem{DJ}
M. Doran and J. Jaeckel, {\rm \prevd} {\bf D66}, 043519 (2002)

\bibitem{carroll}
S. Carroll, {\prlet} {\bf 81} 3067 (1998)

\bibitem{BS}
M. Bucher and D.N. Spergel, {\em \prevd} {\bf D60}, 043505 (1999)

\bibitem{B1}
B. Carter, in {\it Gravitational Radition}, proceedings of the 1982 Les Houches Summer School, ed N. Deruelle and T. Piran, North Holland, Amsterdam, 1983.

\bibitem{B2}
B. Carter, {\em \prevd} {\bf D7}, 1590 (1973)

\bibitem{BM2}
R.A. Battye and A. Moss, {\em in preparation}
      
\bibitem{vil}
A. Vilenkin, {\em \prlet} {\bf 53}. 1016 (1994)

\bibitem{kib}
T.W.B. Kibble, {\em \prevd} {\bf 33}, 328 (1986)

\bibitem{bbs}
R.A. Battye, M. Bucher and D.N. Spergel, astro-ph/9908047

\bibitem{fried}
A. Friedland, H. Murayama and M. Perelstein, {\em \prevd} {\bf D67}, 043519 (2003) 

\bibitem{BCCM}
R.A. Battye, B. Carter, E. Chachoua and A. Moss, hep-th/0501244

\bibitem{CDS}
R.R. Caldwell, R. Dave and P.J. Steinhardt, {\em \prlet} {\bf 80}, 1582

\bibitem{WL}
J. Weller and A. Lewis, {\em MNRAS.} {\bf 346}, 987 (2003)

\bibitem{BD}
R. Bean and O. Dore, {\em \prevd} {\bf D69}, 083503 (2004)

\bibitem{CAMB}
A. Lewis, A. Challinor and A. Lasenby, {\em \jast} {\bf 538}, 473 (2000)

\bibitem{SZ}
U. Seljak and M. Zaldarriaga, {\em \jast} {\bf 469}, 437 (1996)

\bibitem{MCMS}
L. Conversi, A. Melchiorri, L. Mersini and J. Silk, {\em Astropart. Phys.} {\bf 21}, 443

\bibitem{LB}
A. Lewis and S.L. Bridle, {\em \prevd} {\bf D66}, 103511 (2002)

\bibitem{Bennett}
C.L. Bennett et al, {\em Ap. J. Suppl.} {\bf 148}, 1 (2003)

\bibitem{verde}
L. Verde et al, {\em Ap. J. Suppl.} {\bf 148}, 195 (2003)

\bibitem{VSA}
C. Dickinson et al, {\em MNRAS} {\bf 353}, 732 (2004)

\bibitem{CBI}
A.C.S. Readhead et al, {\em \jast} {\bf 609}, 498

\bibitem{acbar}
C.L. Kuo et al, {\em \jast} {\bf 600}, 32 (2004)

\bibitem{perc}
W. Percival et al, {\em MNRAS} {\bf 337}, 1068 (2002) 

\bibitem{perm}
S. Perlmutter et al, {\em \jast} {\bf 517}, 565 (1999)

\bibitem{hamuy}
M. Hamuy et al, {\em Astronomical J.} {\bf 112}, 2391 (1996)

\bibitem{hst}
W.L. Freedman et al, {\em \jast} {\bf 553}, 47 (2001)

\bibitem{bbn}
S. Burles, K.M. Nollet and M.S. Turner, {\em \jast} {\bf 552}, L1 (2001)

\bibitem{dyn}
J.L. Tinker, D.H. Weinberg, Z. Zheng, I. Zehavi, astro-ph/0411777

\bibitem{WA1}
J. Weller and A. Albrecht, {\em \prlet} {\bf 86}, 1939 (2001)

\bibitem{WA2}
J. Weller and A. Albrecht, {\em \prevd} {\bf D65}, 103512 (2002)

\bibitem{Huey}
G. Huey et al {\em \prevd} {\bf D59}, 063005 (1999)

\bibitem{HHM}
Z. Haiman, G. Holder and J.J. Mohr, {\em \jast} {\bf 553}, 545 (2000)

\bibitem{BW2}
R.A. Battye and J. Weller, {\em \prevd} {\bf D68}, 083506 (2003)

\bibitem{HE}
D.J. Eisenstein, in {\em Wide-field Multi-Object spectroscopy}, ASP Conference series, ed. A. Dey

\bibitem{CT}
R. Crittenden and N. Turok, {\em \prlet} {\bf 76}, 575 (1996)

\bibitem{wkl1}
D. Huterer, {\em prevd} {\bf D65}, 063001 (2002)

\bibitem{BC}
S. Boughn and R. Crittenden, {\rm Nature} {\bf 427}, 45 (2004)

\bibitem{ein}
D.J. Eisenstein et al, astro-ph/0501171

\bibitem{aic}
H. Akaike, {\em IEEE Trans. Auto. Control} {\bf 19}, 716 (1974)

\bibitem{bic}
G. Schwarz, {\em Annals of Statistics} {\bf 5}, 461 (1978)

\bibitem{liddle}
A.R. Liddle, {\em MNRAS} {\bf 351}, L49 (2004)

\bibitem{teg}
A. de. Oliveria-Costa, M. Tegmark, M. Zaldarriaga, A. Hamilton, {\em \prevd} {\bf D69}, 063516 (2004)

\end{thebibliography}
\end{document}